\documentstyle[epsf,prl,aps,multicol]{revtex}
\begin{document}
\draft 

\title{A Search for Time Variation of the Fine Structure
Constant}

\author{John K. Webb$^1$, Victor V. Flambaum$^1$,
Christopher W. Churchill$^2$, Michael J. Drinkwater$^1$, John
D. Barrow$^3$} 

\address{$^1$School of Physics, University of New South Wales, Sydney,
NSW 2052, Australia} 

\address{$^2$Department of Astronomy \& Astrophysics, Pennsylvania\\
State University, University Park, PA, 16802, USA}

\address{$^3$Astronomy Centre, University of Sussex, Brighton, BN1 9QJ, UK}

\date{Accepted 1998 December 11 for publication in {\em Physical Review
Letters}} 

\maketitle

\begin{abstract}
A method offering an order of magnitude sensitivity gain is described
for using quasar spectra to investigate possible time or space
variation in the fine structure constant $\alpha$. Applying the
technique to a sample of 30 absorption systems, spanning redshifts
$0.5<z<1.6$, obtained with the Keck I telescope, we derive limits
on variations in $\alpha$ over a wide range of epochs.  For the whole
sample $\Delta \alpha /\alpha =-1.1\pm 0.4\times 10^{-5}$.  This
deviation is dominated by measurements at $z>1$, where $\Delta \alpha
/\alpha = -1.9\pm 0.5\times 10^{-5}$. For $z<1$, $\Delta \alpha
/\alpha = -0.2\pm 0.4\times 10^{-5}$, consistent with other known
constraints.  Whilst these results are consistent with a time-varying
$\alpha$, further work is required to explore possible systematic
errors in the data, although careful searches have so far not revealed
any.

\end{abstract}

\begin{multicols}{2}
\narrowtext

There are several theoretical motivations to search for space-time
variations in the fine structure `constant', $\alpha$.  Theories which
attempt to unify gravity and other fundamental forces may require the
existence of additional compact space dimensions. Any cosmological
evolution in the mean scale factor of these additional dimensions will
manifest itself as a time--variation of our bare three-dimensional
coupling `constants', \cite{Marciano}.  Alternatively, theories have
been considered which introduce new scalar fields whose couplings with
the Maxwell scalar $F_{ab}F^{ab}$ allow time--varying $\alpha$
\cite{carroll}.  The measurement of any variation in $\alpha$ would
clearly have profound implications for our understanding of
fundamental physics.

Spectroscopic observations of gas clouds seen in absorption against
background quasars can be used to search for time variation of
$\alpha$.  Analyses involving optical spectroscopy of quasar absorbers
have concentrated on the relativistic fine-structure splitting of
alkali-type doublets; the separation between lines in one multiplet is
proportional to $\alpha^2$, so small variations in the separation are
directly proportional to $\alpha$, to a good approximation.

Whilst the simplicity of that method is appealing, the relativistic
effect causing the fine splitting is small, restricting the potential
accuracy. We demonstrate below how a substantial sensitivity gain is
achieved by comparing the wavelengths of lines from {\it different}
species, and develop a new procedure, simultaneously analysing the
MgII~2796/2803 doublet and up to five FeII transitions, (FeII~2344,
2374, 2383, 2587, 2600\AA), from three different multiplets. These
particular transitions are chosen because: {\rm (i)} they are commonly
seen in quasar absorption systems, {\rm (ii)} they fall into and span
a suitable rest-wavelength range, {\rm (iii)} an excellent database
was available \cite{Churchill}, {\rm (iv)} extremely precise
laboratory wavelengths have been measured, {\rm (v)} the large Fe and
Mg nuclear charge difference yields a considerable sensitivity gain.

We describe the details of the theoretical developments in a separate
paper \cite{Dzuba}, here summarizing the main points. The energy
equation for a transition from the ground state within a particular
multiplet, observed at some redshift $z$, is given by
\begin{eqnarray}
\begin{array}{ll}
E_z= & E_c+Q_1Z^2\left[ \left( \frac{\alpha _z}{\alpha _0}\right)^2
-1\right] + \\ 
& K_1({\bf LS})Z^2\left( \frac{\alpha _z}{\alpha _0}\right)^2 
+ K_2({\bf LS})^2Z^4\left( \frac{\alpha _z}{\alpha _0}\right)^4
\label{eq1}
\end{array}
\end{eqnarray}
where $Z$ is the nuclear charge, {\bf L} and {\bf S} are the electron
total orbital angular momentum and total spin respectively, and $E_c$
is the energy of the configuration centre. The term in the coefficient
$Q_1$ describes a relativistic correction to $E_c$ for a given change
in $\alpha$, $\alpha _0$ is the zero redshift value, and $\alpha_z$ 
is the value at some redshift $z$. Re-arranged, this gives
\begin{eqnarray}
\begin{array}{ll}
E_z= & E_{z=0}+\left[ Q_1+K_1({\bf LS})\right] Z^2\left[ \left( \frac{\alpha
_z}{\alpha _0}\right) ^2-1\right] + \\ 
& K_2({\bf LS})^2Z^4\left[ \left( 
\frac{\alpha _z}{\alpha _0}\right)^4-1 \right] 
\label{eq2}
\end{array}
\end{eqnarray}
Eq.(\ref{eq2}) is an extremely convenient formulation, the second and
third terms contributing only if $\alpha$ deviates from the
laboratory value.  Accurate values for the relativistic coefficients,
$Q_1,K_1$ and $K_2$, have been computed using relativistic many-body
calculations and experimental data. The coefficients and laboratory
rest wavelengths are given in eq.(\ref {MgFe}). For FeII, the
relativistic coefficients, ($Q_1$), are at least one order of
magnitude larger than the spin-orbit coefficients, ($K_1$). The
variation of the FeII transition frequencies with $\alpha $ is thus
completely dominated by the $Q_1$ term. In MgII, the relativistic
corrections are small due to the smaller nuclear charge $Z$ (see
eq.(\ref {eq2})), so whilst a change in $\alpha$ induces a relatively
large change in the observed wavelengths of the FeII transitions, the
change is small for MgII. The relative shifts are substantially
greater than those for a single alkali doublet alone (such as MgII),
so MgII acts as an `anchor' against which the larger FeII shifts are
measured. A comparison of the observed wavelengths of light and heavy
atoms thus provides a dramatic increase in sensitivity compared to
analyses of alkali doublets alone.

Since it is already clear from previous observational constraints that
any change in $\alpha$ will be very small \cite{Varsh}, \cite{Cowie},
\cite {Drinkwater}, it is vital that $E_{z=0}$ is known accurately
enough.  Indeed, the {\it change} in the frequency interval between
MgII~2796 and FeII~2383 induced by a fractional change $\Delta \alpha
/\alpha =10^{-5}$ is, using eq.(\ref{MgFe}), 0.03 cm$^{-1}$. Thus,
independent of the quality of the observations, the limiting accuracy
in a determination of $\Delta \alpha /\alpha$ is $\sim 10^{-5}$ for an
uncertainty in the laboratory frequency of $\sim 0.03$ cm$^{-1}$. This
highlights the advantage of comparing light and heavy atoms. Previous
analyses of alkali doublets, \cite{Varsh}, \cite {Cowie}, have used
frequencies of about the accuracy above, but have been restricted to
placing limits of $\Delta \alpha /\alpha \approx 10^{-4}$.

Very precise laboratory spectra of the MgII~2796 and MgII~2803 lines
have recently been obtained \cite{Pickering}, in excellent agreement
with previous accurate measurements of MgII~2796 alone
\cite{Drullinger}, \cite{Nagourney}.  Similarly precise FeII
hollow--cathode spectra exist \cite{Nave}.  Inserting these laboratory
wavelengths and our $Q$ and $K$ coefficients into eq.(\ref{eq2}), we
obtain the dependence of frequency on $\alpha$ for MgII (top two
equations in eq.(\ref{MgFe}) below) and FeII:

\begin{eqnarray}
\begin{array}{rl}
^2P~~J=1/2:~ & \omega =35669.286(2)+119.6x \\ 
~~J=3/2:~ & \omega =35760.835(2)+211.2x \\ 
^6D~~J=9/2:~ & \omega =38458.9871(20)+1394x+38y \\ 
~~J=7/2:~ & \omega =38660.0494(20)+1632x+0y \\ 
^6F~~J=11/2:~ & \omega =41968.0642(20)+1622x+3y \\
~~J=9/2:~ & \omega =42114.8329(20)+1772x+0y \\ 
^6P~~J=7/2:~ & \omega =42658.2404(20)+1398x-13y \\ 
\end{array}
\label{MgFe}
\end{eqnarray}
where $x=[(\frac{\alpha _z}{\alpha _0})^2-1]$ and $y=[(\frac{\alpha _z}{%
\alpha _0})^4-1]$.

The astronomical data used for this analysis was obtained using the
HIRES echelle spectrograph \cite{Vogt} on the Keck I 10m telescope
during three observing runs in 1994-1996.  High-quality spectra of 25
quasars were obtained, in which intervening absorption systems at
low/intermediate redshift have been identified containing FeII, MgII,
and other species. Full observational details are given in
ref. \cite{Churchill}.


We now determine the relative positions of the FeII and MgII lines and
estimate $\Delta \alpha /\alpha$ for each absorption system in the
sample.  Measuring each line (ie. MgII~2796, MgII~2803 and 5 FeII
lines) independently is not optimal, because the number of fitting
parameters is not minimised, as discussed below. The procedure used is
iterative, where all available lines are fitted simultaneously with
Voigt profiles, using VPFIT, a non-linear least-squares programme
designed specifically for analysing quasar absorption spectra
\cite{Webb87}.  We minimise the total number of free parameters by
linking physically related parameters: (i) the redshifts of the
corresponding FeII and MgII components are tied; (ii) the column
densities N(FeII), N(MgII) for individual components can vary
independently of each other, but the velocity dispersion parameters
$b$(FeII), $b$(MgII), ($b=\sqrt{2}\sigma$), in corresponding components,
are constrained by $\sqrt{(24/56)}b$(MgII)$\leq b$(FeII)$\leq
b$(MgII).  We carry out 150 separate fits to each absorption system,
varying $\alpha$ slightly each time, using eq.(\ref{MgFe}) to compute
the input rest-frame wavelengths.  That procedure is performed 3
times; twice where the $b$ parameters are related according to the two
extremes above (ie. thermal or turbulent line-broadening) and a third
time where all $b$ parameters vary independently.  The fitting
procedure returns a value of $\chi _{min}^2$ which was computed as a
function of $\Delta \alpha /\alpha$. We used the standard statistical
procedure of estimating 1$\sigma$ errors on $\Delta \alpha /\alpha$
from $\chi_{min}^2\pm 1$.

Several consistency checks are imposed before accepting a result.
First, the $\chi _{min}^2$ for each individual spectral region being
fitted must be statistically acceptable (ie. its reduced $\chi ^2$ is
$\approx 1$). It follows that $\chi _{min}^2$ for the fit as a whole
is statistically consistent with the number of degrees of freedom for
that fit. Second, we require statistical consistency between the three
separate analyses (for the three different $b$ constraints). If any of
the three $\Delta \alpha /\alpha$ differ by more than 1$\sigma$ from
either of the other two, the system is rejected. These criteria lead
to the rejection of only about 1/10 of the sample, reflecting the good
general statistical robustness of the procedure.  The final $\Delta
\alpha /\alpha$ accepted corresponded to the lowest of the 3 values of
$\chi _{min}^2$.

The best-fit values of $\Delta \alpha /\alpha$, and the 1$\sigma$
error bars, are plotted against redshift in Fig. 1.  For $z<1$ there
is no departure from the present-day value.  The scatter
in the data is consistent with the individual error bars, {\it ie.}
there is no evidence for any space or time--variations in $\alpha$.
However, at $z>1$ the situation is less clear, with 14 points
giving $\chi^2 = 34.9$, 3 falling above zero and the rest below.
Fig. 1 shows there is a dip in a relatively narrow redshift interval
$0.9<z<1.2$, where 9 out of the 10 points lie below zero, and it is
this dip which cause the overall $z>1$ $\chi^2$ to be high.  Assuming
that this is a statistical fluctuation and that we have somehow
underestimated the errors, we can increase the error
bars on each point by a constant amount, $S$, where $S \geq 0$ such
that the reduced $\chi^2 = 1$, {\it ie.} the error on the i$^{th}$
point becomes $\sigma'_i = \sigma_i + S$ (where $\sigma_i$ is the value
illustrated in Fig. 1).  The results of this procedure, for the sample
as a whole, and in two redshift ranges are:
\smallskip
\begin{equation}
\begin{array}{ll}
\smallskip \frac{\Delta \alpha }\alpha =-1.09\pm 0.36\times 10^{-5} & %
\hspace{0.3in}(0.6<z<1.6) \\
\smallskip \frac{\Delta \alpha }\alpha =-0.17\pm 0.39\times 10^{-5} & %
\hspace{0.3in}(0.6<z<1.0) \\
\smallskip \frac{\Delta \alpha }\alpha =-1.88\pm 0.53\times 10^{-5} & %
\hspace{0.3in}(1.0<z<1.6)
\end{array}
\label{results}
\end{equation}
The values of $S$ are, in order, $S=0.46, 0.06, 0.53$.  The whole
sample departs from zero at the $3.0\sigma$ level towards smaller
values of $\alpha$.  However, the $z<1$ points alone show no
significant trend, suggesting that there are no significant errors in
the adopted laboratory FeII and MgII wavelengths, or in the general
procedure. However, at $z>1$, there is a $3.5\sigma$ deviation.  We
have also experimented with other statistical methods which also give
more conservative error estimates than a simple weighted mean ({\it
eg.} a Bayesian method), and obtain results consistent with the above
(to be reported elsewhere).


What systematic (non-physical) effects could mimic such an effect
(either a general trend towards negative $\Delta \alpha /\alpha$ for
$z>1$, or, the curious `cusp' in the range $0.9<z<1.2$)? The quality
of the spectroscopic data does not deteriorate with redshift (this
would be revealed by larger error bars at higher redshift) but in fact
improves \cite{Churchill}. The unmodified error bars reflect
uncertainties due to signal-to-noise, spectral resolution, velocity
structure, the number of absorption lines fitted, and assume we have
properly de-convolved the absorption features into their correct
number of individual components.  However this assumption may be
wrong.  There may be weak blended unresolved lines, even with data of
the high quality we have here.  For a specific absorption system, an
interloper could indeed mimic a shift, {\it eg.} if it were blended
with one of the MgII~2796 or 2803 `anchors'.  Nevertheless it is
unlikely that this occurs preferentially for $z>1$ and not for $z<1$.
Nor should {\it random} blending create a negative $\Delta \alpha
/\alpha$ trend; it should merely increase the scatter about a mean of
zero.  An extensive search for a {\it systematic} blend ({\it ie.} a
weak transition from some other species at the same redshift as the
absorbing cloud) falling close to one of the Mg or Fe absorption
lines, proved negative (details to be reported elsewhere).

Uncertainties in the Fe or Mg laboratory wavelengths cannot be
responsible, since any errors would be an order of magnitude below the
observed effect, and in any case, the $z<1$ points reveal no
offset. We also checked to see whether uncertainty in the instrumental
resolution could introduce errors, and found the results insensitive
to this.  It is conceivable that certain absorption
redshifts could be less reliable, due to the positioning of the FeII
and MgII lines with respect to the echelle order edges, where
wavelength calibration may be worse.  At higher wavelengths, the HIRES
optical format is such that there are gaps in the wavelength
coverage. Also, the ThAr calibration spectrum has fewer lines per unit
wavelength at higher wavelengths.  These effects could mean that
wavelength calibration is less reliable for higher redshift points
than the lower redshift ones.  Alternatively, perhaps there is a
subtle bug in the wavelength calibration software. The $z>1$ effect we
have found is very small: a gradual drift of $\sim 2.5$ times the
wavelength calibration residuals, over the range corresponding to the
observed FeII and MgII lines, could produce an effect of the
significance we have found.  Previous analyses of similar quality data
have not demanded such high precision, so such errors could have gone
unnoticed in other spectroscopic analyses.

Can any of these effects produce the apparent trend we find?  We have
carried out extensive numerical experiments aimed specifically at
testing for these potential systematic problems.  These involved
carrying out an identical analysis on the wavelength calibration
spectra (ThAr) as for the QSO spectra, measuring the emission line
wavelengths and inferring a `change' in $\alpha$ from shifts betwen
the measured and laboratory wavelengths (details to be reported
elsewhere).  The results were unambiguous, in that any errors in the
wavelength calibration across all spectral orders are so small that
they could not contribute significantly to the apparent offset at $z >
1$, unless the literature values for the calibration spectral lines
(ThAr) are substantially in error, with a gradual non-linear shift in
the ThAr wavelengths.

We consider all of the above potential errors to be rather unlikely.
Since the effect is dominated by the $0.9 < z < 1.2$ points, and
because a genuine physical effect confined to one specific epoch in
the history of the universe does not seem at present to be well
motivated by theoretical expectations, we presume the explanation is
that there are additional undiscovered velocity components in the
absorbing gas for those particular absorption sytems, even though all
points have been subjected to the same analysis.  Further observations
of different species in these clouds could answer this.  Our results
should thus be regarded as stringent upper limits on any possible
time-variation rather than a positive detection of a change.

We may compare our results with other recent values.  Observations at
$z\sim 3$ \cite{Cowie} have yielded an upper limit $\left| \Delta
\alpha /\alpha \right| <3.5\times 10^{-4}$. Our analysis is at lower
redshift, so a comparison requires choosing some (arbitrary)
functional form of the evolution. At lower redshifts, a recent
analysis of radio wavelength spectra of atomic hydrogen and molecular
gas \cite{Drinkwater} gives limits $\left| \Delta \alpha /\alpha
\right| <3\times 10^{-6}$ at two redshifts, $z=0.25,0.68$.  An
additional recent constraint of $3.5 \pm 5.5 \times 10^{-6}$
\cite{Cowie} comes from a comparison of hyperfine and optical
redshifts.  These limits are consistent with the results given in
eq.(\ref{results}). The strongest terrestrial constraint on the time
evolution of $\alpha$ comes from the Oklo natural nuclear reactor
\cite {Shylakhter}. The Oklo event is estimated to have taken place
around 1.8 billion years ago (corresponding to $z \approx 0.1$).  We
adopt a cosmological model with $q_0=1/2$, $\Lambda =0$ and take the
age of the universe to be $13 \times 10^9$ years.  The Oklo data has
recently been re-examined \cite{Damour}, and upper limits on a change
in $\alpha$ are $-0.9\times 10^{-7}<(\alpha
^{Oklo}-\alpha_0)/\alpha_0<1.2\times 10^{-7}$, assuming the weak and
strong couplings are unchanged.  These bounds are clearly consistent
with our results for $z<1$. For the cosmological parameters quoted, if
we adopt $\bar z=0.8$, the mid-point of our lower redshift range,
eq.(\ref{results}) implies
\begin{equation}
\begin{array}{ll}
\frac{\left\langle \dot \alpha \right\rangle }\alpha =-2.2\pm 5.1\times
10^{-16} ~yr^{-1}\hspace{0.2in}(0.6<z<1.0) &  \\ 
\end{array}
\label{timederiv}
\end{equation}

What other physical phenomena, other than time variability of
$\alpha$, could give rise to the observational effect we report? The
spacing of the MgII and FeII isotopes is such that a significant
change in the isotopic ratios could explain the observations. However,
the change would need to be substantial; for example, this would
require {\it most} of the Mg in the universe at $z>1$ to be in
$^{26}$Mg (the present epoch abundance is $\sim 10\%$), and a physical
mechanism found to convert almost all the $^{26}$Mg into $^{24}$Mg by
the present epoch.  If large scale magnetic fields exist, and the
quasar light is polarised, these could potentially give rise to
correlated apparent shifts in absorbers in neighbouring regions of the
universe.  However, for magnetic fields to be responsible for the
global effect in $\alpha$ for $z > 1$ (but not for $z < 1$), a sharp
variation at $z\sim 1$ or some form of oscillatory variation would be
required, both of which are hard to motivate.

Some authors have suggested more exotic forms of evolution of the
constants, including oscillations \cite{Marciano}.  These could arise
from new light bosons with mass $m$, producing periodic variations in
the frequency of the radiation emitted at high $z$, with a modulation
frequency $\sim m^{-1}$ \cite{hill}.  The creation of time--varying
$\alpha$ by approximate global symmetry (\cite{carroll}) allows
oscillatory variations introduced by the decaying mean oscillations of
a scalar field coupling to $F_{ab}F^{ab}$.  We note that as we move
above $z=1.25$, for a critical-density non-inflationary universe, we
can encounter causally disconnected regions of the universe. The
observational sensitivity achieved by these observations exceeds that
of current microwave background observations and a larger dataset than
ours may contain important new cosmological information.

The work we have presented here demonstrates the possibilites for
extending this type of study to incorporate different species, other
than Mg and Fe, spanning a wider redshift baseline, so equally
impressive constraints should be obtainable at higher redshifts.
Future analyses of other species will be hampered by the lack of
accurate laboratory wavelengths. We hope that this paper will provide
an impetus for new high-precision laboratory measurements.

\bigskip

We are very grateful to R. Carswell, J. Charlton, V. Dzuba,
A. Fernandez-Soto, J. Garcia-Bellido, R. Learner, C. Lineweaver,
J. Magueijo, D. Morton, M. Murphy, J. Pickering, O. Sushkov,
A. Thorne, A. Vidal-Madjar, S. Vogt and the HIRES team, D. Wineland
for various important contributions.  We also thank A. Dryer and SUN
Microsytems Australia Pty Ltd for computers and L. Evans for computing
assistance.  JDB is supported by PPARC.

\end{multicols}

\begin{figure}
\epsfxsize=19cm \epsffile{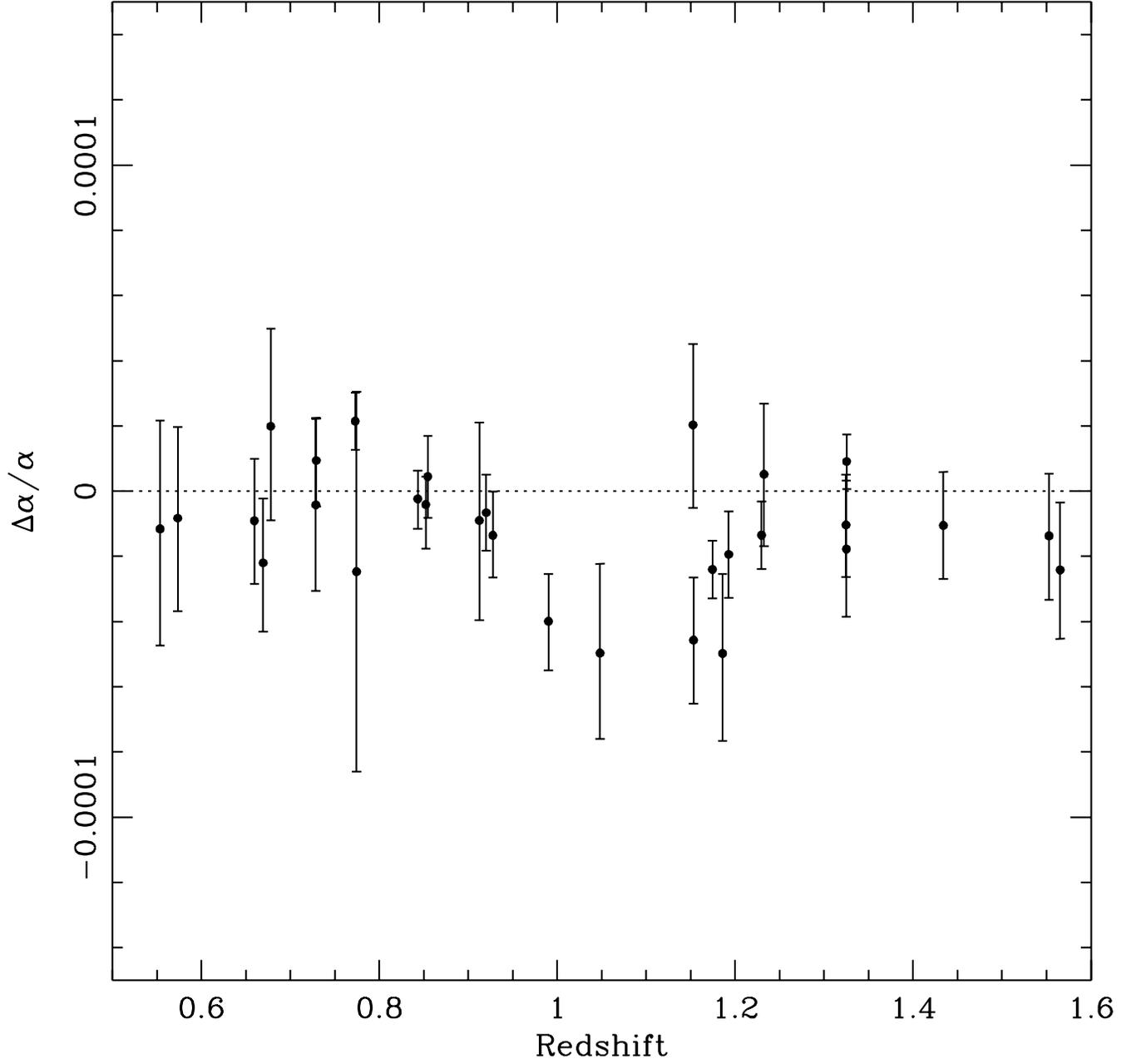}
\caption{Plot of $\Delta \alpha / \alpha$ vs $z$ for
30 FeII/MgII absorption systems.  All the transitions given in equation 3
were used, where available.  Plotted error bars were determined from
$\chi^2_{min} \pm 1$ (but statistical results estimated using larger
errors, as discussed in the text).}
\end{figure}
\end{document}